\documentclass[aps,prl,nofootinbib,twocolumn,superscriptaddress,preprintnumbers]{revtex4}

\usepackage{amssymb}
\usepackage{amsmath}
\usepackage{epsfig}
\usepackage{hyperref}
\usepackage{breakurl}

\makeatletter
\def\simgt{\mathrel{\lower2.5pt\vbox{\lineskip=0pt\baselineskip=0pt
           \hbox{$>$}\hbox{$\sim$}}}}
\def\simlt{\mathrel{\lower2.5pt\vbox{\lineskip=0pt\baselineskip=0pt
           \hbox{$<$}\hbox{$\sim$}}}}
\makeatother

\def\mysection#1{{\bf #1.} }

\newcommand{\be}{\begin{equation}}
\newcommand{\ee}{\end{equation}}
\newcommand{\bea}{\begin{eqnarray}}
\newcommand{\eea}{\end{eqnarray}}
\newcommand{\beq}{\begin{eqnarray}}
\newcommand{\eeq}{\end{eqnarray}}

\newcommand{\no}{\nonumber}

\def\mysection#1{{\bf #1.} }

\def\lsim{\mathrel{\rlap{\lower4pt\hbox{\hskip1pt$\sim$}}
     \raise1pt\hbox{$<$}}}         %less than or approx. symbol
\def\gsim{\mathrel{\rlap{\lower4pt\hbox{\hskip1pt$\sim$}}
     \raise1pt\hbox{$>$}}}         %greater than or approx. symbol

\begin{document}

\title{Superconducting Detectors for Super Light Dark Matter}

\author{Yonit Hochberg}
\affiliation{Theoretical Physics Group, Lawrence Berkeley National Laboratory, Berkeley, CA 94720 \\ Berkeley Center for Theoretical Physics, University of California, Berkeley, CA 94720}
\author{Yue Zhao}
\affiliation{Stanford Institute for Theoretical Physics, Department
of Physics, Stanford University, Stanford, CA 94305, U.S.A.}
\author{Kathryn M. Zurek}
\affiliation{Theoretical Physics Group, Lawrence Berkeley National Laboratory, Berkeley, CA 94720 \\ Berkeley Center for Theoretical Physics, University of California, Berkeley, CA 94720}

\begin{abstract}

We propose and study a new class of of superconducting detectors which are sensitive to ${\cal O}({\rm meV})$ electron recoils from dark matter-electron scattering.  Such devices could detect dark matter as light as the warm dark matter limit, $m_X \gsim 1 \mbox{ keV}$.  We compute the rate of dark matter scattering off of free electrons in a (superconducting) metal, including the relevant Pauli blocking factors.  We demonstrate that classes of dark matter consistent with terrestrial and cosmological/astrophysical constraints could be detected by such detectors with a moderate size exposure.
\end{abstract}

\maketitle

%%%%%%%%%%%%%
\mysection{Introduction}
\label{sec:intro}
The search for the identity of dark matter (DM) is in an exciting
and rapidly developing era.  Theories of weakly interacting massive particles (WIMPs) for DM, being predictive and testable, have been the primary focus of both theory and experiment for the last thirty years. Strong constraints from direct detection experiments, such as Xenon100~\cite{Aprile:2012nq}, LUX~\cite{Akerib:2013tjd} and SuperCDMS~\cite{Agnese:2014aze}, along with the absence of new physics signals from the LHC, have, however, been painting such models as increasingly constrained and tuned.   Further, because the energy threshold of direct detection experiments searching for WIMPs is typically 1-10 keV,  these experiments lose sensitivity to DM particles with mass below 10 GeV.   At the same time, DM candidates with low masses are theoretically well-motivated:
asymmetric dark matter~\cite{Kaplan:2009ag,Zurek:2013wia} and strongly interacting massive particles~\cite{Hochberg:2014dra} are examples in which the natural mass scale of the DM sits beneath the $\sim10$~GeV scale.

A new frontier for massive DM thus opens for $1 \mbox{ keV}
\lsim m_X \lsim 10 \mbox{ GeV}$, with the lower bound set
approximately by warm dark matter constraints, {\em e.g.} from
phase space packing~\cite{Tremaine:1979we,Boyarsky:2008ju} or the
Lyman-$\alpha$ forest~\cite{Boyarsky:2008xj}.  For elastic scattering
processes, the deposited energy is $E_D \simeq q^2/(2 m_{e,N})$,
where $q \sim \mu_r v_X$ is the momentum transfer with $v_X\sim 10^{-3}$ the incoming DM
velocity, $\mu_r$ the reduced mass of the system and $m_{e,N}$ is the mass of the
target electron or nucleus $N$.  Thus for 100~MeV DM, an eV of energy is
deposited for scattering off a nucleus.  Inelastic processes, such as electron ionization or excitation above
a band gap, may occur when the DM kinetic energy
exceeds the binding energy.  Utilizing a semi-conducting crystal
such as germanium, with a band gap of 0.7~eV, implies potential sensitivity to DM as light as ${\cal O}({\rm MeV})$~\cite{Essig:2011nj,Graham:2012su}.  SuperCDMS is already working to lower its threshold to 300 eV \cite{Agnese:2014aze}, constraining 1 GeV mass DM.

To go well below this, as low as the warm DM limit at ${\cal O}({\rm keV})$, requires a
different kind of technology; in this case one must be able to
access electron recoil energies as low as ${\cal O}({\rm meV})$.  The purpose of
this letter is to investigate a proof of principle experiment to
search for DM down to the warm DM limit.  Devices utilizing
superconductors, we will show, are ideal for this purpose, as they
can be sensitive to extremely small energy depositions.  In fact, in
cold metals, the limit on the sensitivity of the experiment to low
energy DM recoils is set by the ability to control the noise
rather than by an inherent energy gap in the detector.

The targets we discuss are metals, with the DM interacting with free
electrons in the Fermi sea.  The DM scattering rate is limited by
Pauli blocking for electrons locked deep in the sea, yielding a
suppression factor of order the energy transfer over the Fermi
energy; the suppression is, {\it e.g.}, of order $\sim10^{-4}$ for a
DM-electron scattering with meV energy deposition in a typical metal
such as aluminum. As we will show, DM models satisfying all
astrophysical and terrestrial constraints are detectable despite the
Pauli blocking effect, extending the conceptual reach of the
detection method down to DM masses of ${\cal O}({\rm keV})$.

%%%%%%%%%%%%%%%%
\mysection{Detection with Superconductors}
The challenge in designing a detector to observe DM with low energy deposits is to achieve a large target mass, while keeping
noise low. Detection of small energy depositions is by now
well-established; superconductors, with a meV
superconducting-gap, have sensitivity to energies at this scale.
Transition edge sensors (TES) and Microwave Kinetic Inductance
Devices (MKIDs) have been utilized to detect microwaves and x-rays
with sub-meV to keV energies in astrophysical applications.  For
example, TESs with sensitivity to energy depositions not very far
from our range of interest already exist: Refs.~\cite{NIST,spica,JPL} have demonstrated
noise equivalent power in the range $\sim 10^{-19}-10^{-20}
\mbox{W}/\sqrt{\mbox{Hz}}$.  This translates to a sensitivity of
$\sim 50 - 300$ meV of energy over a read-out time of $\sim 10 $ ms.  Thus current technology could already start probing new regions of parameter space, though not yet at the ${\cal O}({\rm meV})$ level of sensitivity required for probing down to keV dark matter. Since the energy resolution scales with $\sqrt{T^3 V}$, with $T$ the heat bath temperature and $V$ the TES volume, the required improvement could be made by lowering $T$ (from, for instance, 100~mK to 10~mK) and further decreasing the heat capacitance of the TES by reducing the volume.
%improvement could be made by lowering both the heat bath temperature
%(from, for instance, 100 mK to 10 mK) and decreasing the heat
%capacitance of the TES by reducing the volume.

The TES and MKID, however, have very low masses---an MKID is
typically a nano-gram in weight, while TESs are approximately 50 microns on
a side and a fraction of a micron thick. As a result, they do not make good detectors
themselves.  Their masses cannot simply be increased since this would decrease their sensitivity to small deposits of energy.  An
alternative is then to use the TES or the MKID merely as heat sensors which
register small deposits of energy from a much larger target mass, an
`absorber.'

For the absorber, we choose a supeconductor; a superconductor features an energy gap which controls the thermal noise in the absorber.
As a DM particle hits a free electron in the Fermi sea of the
absorber, %if the absorber itself is a superconductor,
the recoiling electron will deposit an ${\cal O}(1)$ fraction of its energy into
breaking Cooper pairs, creating quasiparticles in the
superconductor.  These quasiparticles random walk in the superconductor until the energy stored in them can be collected.  Two possibilities for the collection are that the quasiparticles (I) re-combine and create an
athermal phonon or (II) are absorbed on collection fins on the surface of the absorber.
%These quasiparticles random walk in the
%superconductor until they either (I) re-combine and create an
%athermal phonon or (II) are absorbed.

In the former case, the athermal phonon may break Cooper pairs in the MKID, leading to an observable change in the kinetic
inductance.  In the latter case, the quasiparticles may reach a
collection fin on the surface of the absorber.  The fins should have a lower gap than the absorber, both to control noise and to facilitate collection of energy into the fins. %; the absorber should then have a higher gap than the fins, to control the noise.
%(with the absorber having a higher gap than the fins, to control noise).
%(with the absorber $T_c$ above that of the fins, to control noise).
The collection fins are connected to the TES which registers the heat.
The quasiparticle lifetimes are sufficiently long and their velocities sufficiently high that even if the collection fin area on the absorber is small, the quasiparticles ricochet sufficiently many times that they are very efficiently channeled from the absorber into the collection fins and on to the TES. Aluminum is an example of an ultra-pure metal that makes for a good absorber: with quasiparticle lifetimes of order a millisecond~\cite{0811.1961} and velocities of order the Fermi velocity $v_F\sim 10^{-2}c$, its gap of $\sim0.3\; {\rm meV}$ pairs well with gapless gold collection fins. We note that the scattering length in the absorber sets the upper bound on its unit size% the size of a unit absorber
---of order $\sim 5$~mm in ultra-pure aluminum---such that many small absorbers must be multiplexed for large exposure.

%The quasiparticle lifetimes are sufficiently long (well in excess of $\mu$s in {\it e.g.} aluminum) and their velocities sufficiently high (up to the Fermi velocity, $\sim 10^{-2} c$) that even if the collection fin area on the absorber is small, the quasiparticles ricochet sufficiently many times that they are very efficiently channeled from the ({\it e.g.} aluminum, with $\sim 0.3$~meV gap) absorber into the ({\it e.g.} gapless gold) collection fins and on to the TES. (The scattering length of the absorber sets the upper bound on the size of a unit absorber, of order $\sim 5$~mm in ultra-pure aluminum, such that many small absorbers must be multiplexed for large exposure.)

In either case, the MKID or the TES is acting as a calorimeter for the energy
deposited in the absorber.  The underlying design principle sketched
here is of concentration: one seeks to store the deposited energy
non-thermally, whether through quasiparticles or athermal phonons,
and then concentrate them through a collection mechanism onto the
MKID or TES.  This process must happen fairly rapidly, on the
timescale of a millisecond.

Our purpose here is not to advocate for a particular experimental design, but rather simply
to outline how, through improvements to existing technology, sensitivity to
extremely light DM utilizing superconductors may
be feasible.  (Other techniques, such as the use of superfluid helium~\cite{Hertel}, hold promise as well.) The remainder of this letter focuses on the reach of such an experiment into the parameter space of light dark matter.

%%%%%%%%%%%%%%%%%
\mysection{Rates and Backgrounds}\label{sec:rates}
Detection via TESs (or MKIDs) operates by DM scattering off of free electrons
in a metal.  In a superconductor, the free electrons are bound into Cooper pairs, which typically have $\sim{\rm meV}$ (or less) binding energy.
Once the energy in the scattering exceeds this superconducting gap, however, the scattering rate is computed by the interaction with free electrons, times a coherence factor. This factor is ${\cal O}(1)$ for energies just above threshold, and goes to unity for energies above the gap, see {\it e.g.} Ref.~\cite{Tinkham}. In the setups we consider, the gap is below the noise-limited energy resolution, and the coherence factor can be neglected.
%Once the energy in the scattering exceeds this superconducting gap, however, the scattering rate is computed by the interaction with free electrons (times a coherence factor, which yields an ${\cal O}(1)$ factor just above threshold, and goes to unity for energies above the gap, see {\it e.g.} Ref.~\cite{Tinkham}).
The electrons are then described by a Fermi-Dirac
distribution at low temperature.  The typical Fermi energy $E_F$ of
these electrons is $p_F^2/(2 m_e) \sim 10$~eV, with $p_F \sim 50
\mbox{ keV}$ in a typical metal such as aluminum. Scattering with a target electron buried in the Fermi sea can break the Cooper pair if the energy transferred in the scattering is enough to pull an electron out of the sea and above the gap.
As a result, with kinetic energy of the
incoming DM approximately $m_X v_X^2 \sim \mbox{ meV} -  \mbox{
keV}$ for keV to GeV DM, Pauli blocking is important for the DM
scattering rate.   We follow the discussion in~\cite{Reddy:1997yr}
to compute the rate correctly, factoring in the Pauli blocking
effect. We denote the 4-momentum of DM initial and final states by
$P_1$ and $P_3$, the initial and final states of the electron by
$P_2$ and $P_4$, and the momentum transfer $q=(E_D, {\bf q})$. The scattering rate can be estimated via
\begin{eqnarray}
 \label{eq:response}
 \langle n_e\sigma v_{\rm rel}\rangle&=&\int\frac{d^3
p_3}{(2\pi)^3}\frac{ \langle|\mathcal {M}|^2\rangle}{16 E_1 E_2 E_3 E_4}\ S(E_D,|{\bf q}|)\,,\\
S(E_D,|{\bf q}|)&=&2\int\frac{d^3 p_2}{(2\pi)^3}\frac{d^3
p_4}{(2\pi)^3}(2\pi)^4\delta^4(P_1+P_2-P_3-P_4)\nonumber\\
&&\quad \quad \quad \times f_2(E_2)(1-f_4(E_4)),\nonumber
\end{eqnarray}
where $E_D$ is the deposited energy, $ \langle|\mathcal
{M}|^2\rangle $ is the squared scattering matrix element summed and
averaged over spin, and $f_i(E_i)=[
1+\exp(\frac{E_i-\mu_i}{T})]^{-1}$ is the Fermi-Dirac distribution
of the electrons at temperature $T$. $S(E_D,|{\bf q}|)$
characterizes the Pauli blocking effects, and in the limit of
$T\rightarrow 0$, $S(E_D,|{\bf q}|)$ reduces to a simple
Heaviside theta function, with amplitude $m_e^2 E_D/(\pi
|{\bf q}|)$. We perform the integral
numerically in order to capture the entire kinematic range properly.
The total rate (per unit mass per unit time) is then
\begin{eqnarray}
 \label{eq:RateDM}
E_D\frac{dR_{\rm DM}}{d E_D}=\int d v_X f_{\rm MB}(v_X)
E_D\frac{d\langle n_e\sigma v_{\rm rel}\rangle}{d
E_D}\frac{1}{\rho}\frac{\rho_{X}}{m_X}\,.
\end{eqnarray}
Here $\rho$ is the mass density of the detector material, and $\rho_{X} =
0.3\ \textrm{GeV}/\textrm{cm}^3$ the DM mass density. We take the
velocity distribution of the DM $ f_{\rm MB}(v_X)$ to be a modified Maxwell Boltzmann with rms velocity $v_0=220\ \textrm{km}/\textrm{sec}$,
and  cut-off at the escape velocity $v_{\rm esc}=500\ \textrm{km}/\textrm{sec}$.
Since the typical Fermi velocity of a metal is $v_F={\cal
O}(10^3)\;\textrm{km}/\textrm{sec}\gg v_{\rm esc}$, $v_{\rm rel}
\simeq v_F$. The Pauli blocking
effect provides a suppression factor of  order $E_D/E_F$, which we confirm numerically.  An irreducible background is
expected to come from electron-neutrino scattering, which, due to the low energy
deposition in the detector, will be dominated by $pp$
neutrinos~\cite{Bahcall:1987jc,Bahcall:1997eg}. We find that the
solar neutrino background is many orders of magnitude below the
signals we consider, and is hence omitted from further discussion.  We have also checked that backgrounds from Compton scatters (at levels already achieved in an experiment such as CDMS) are not significant.

In what follows we assume that the DM $X$ interacts with electrons
via exchange of a mediator $\phi$.  The generalization of light DM
models will be addressed in future work~\cite{future}; we seek only
to demonstrate proof of principle here. The scattering cross section
between, {\it e.g.}, Dirac DM and free electrons is given by
$\sigma_{\rm scatter} = 16 \pi \alpha_e \alpha_X
\mu_{eX}^2/(m_\phi^2+q^2)^2$, where $\alpha_i\equiv g_i^2/(4\pi)$,
$g_i$ is the coupling of $\phi$ to $i$ with $i=e,X$, $\mu_{eX}$ the
reduced mass of the electron-DM system, and $q$ the momentum
transfer in the process.  This cross-section is related to the
matrix element in Eq.~\eqref{eq:response} via $\sigma_{\rm scatter}
= \frac{ \langle|\mathcal {M}|^2\rangle}{16 \pi E_1 E_2 E_3
E_4}\mu_{eX}^2 $.   We define two related reference cross sections
$\tilde \sigma_{\rm DD}$, corresponding to the light and heavy
mediator regimes:
\beq\label{eq:DD}
\tilde \sigma_{\rm DD}^{\rm light} &=& \frac{16\pi \alpha_e \alpha_X}{q_{{\rm ref}}^4}\mu_{eX}^2\,,\quad q_{\rm ref} \equiv \mu_{eX} v_X\,, \no\\
\tilde \sigma_{\rm DD}^{\rm heavy}&=& \frac{16\pi \alpha_e \alpha_X}{m_\phi^4}\mu_{eX}^2 \,,
\eeq
where $v_X = 10^{-3}$.
The transition between these regimes is set by how large the mediator mass is in comparison to the momentum
transfer. The reference momentum transfer $q_{\rm ref}$ above is chosen for
convenience as a typical momentum exchange.
Note however that for a light mediator, the direct detection cross
section is determined by the minimal momentum transfer in the
process, which is controlled by the energy threshold of the
detector.
%(Note that in the setups we consider, the gap is below the noise-limited energy resolution.)

\begin{figure}[t!]
\begin{center}
\includegraphics[width=0.5\textwidth]{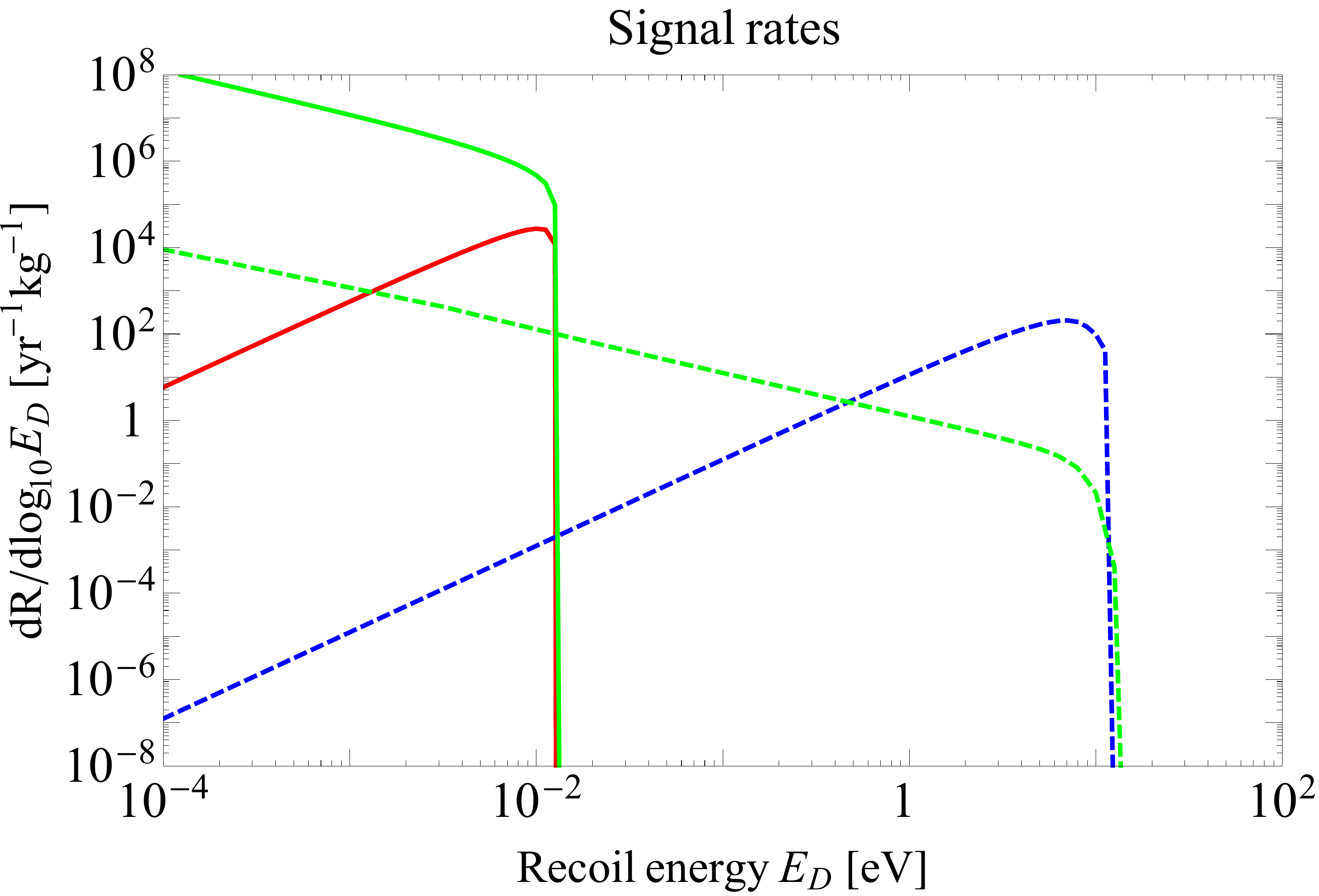}
\caption{Signal rates per kg$\cdot$year, for several benchmark
points of ($m_\phi, m_X, \alpha_X, g_e)$ = $(10\;\mu {\rm eV}, 10\; {\rm keV}, 5\times10^{-14}, 3\times10^{-9})$ [{\bf solid green}],  $(10\; \mu {\rm eV}, 100\; {\rm MeV}, 5\times 10^{-8}, 3\times 10^{-12})$ [{\bf dashed green}], $(1\;{\rm MeV}, 10\;{\rm keV}, 0.1, 3\times 10^{-6})$ [{\bf solid red}], and $(100\;{\rm MeV}, 100\;{\rm MeV}, 0.1, 3\times 10^{-5})$ [{\bf dashed blue}]. We use the Fermi energy of aluminum, $E_F=11.7$~eV.
%The green [red and blue] curves correspond to a particular DM mass along the
%same-colored curve in the left [right] panel of
%Fig.~\ref{fig:DDcomb}.
}\label{fig:rate}
\end{center}
\end{figure}

To establish a notion of the expected number of events, in
Fig.~\ref{fig:rate} we present the differential rate per
kg$\cdot$year as a function of deposited energy for several
benchmark points described in the next section.
When the mediator is effectively massless, namely lighter than the momentum transfer in
the scattering, the rate is peaked at energies near the detector
threshold due to the $1/q^4$ enhancement of the cross section. In contrast, for massive mediators, the rate is peaked at
higher recoil energies.  The reason for the latter behavior is that
as the recoil energy increases, more electrons can be pulled from
deeper in the Fermi sea, resulting in an increased rate.
%When the mediator is
%effectively massless, namely lighter than the momentum transfer in
%the scattering, the rate is peaked at energies near the detector
%threshold. In contrast, for massive mediators, the rate is peaked at
%higher recoil energies.  The reason for the latter behavior is that
%as the recoil energy increases, more electrons can be pulled from
%deeper in the Fermi sea, resulting in an increased rate.
The mass of the mediator determines the scattering distribution in phase
space, but does not control the size of the available phase space. A
cutoff in the differential rate is evident for both light
and heavy mediators, and depends on the DM mass. For heavier DM
(dashed curves), the maximum energy deposition is determined by $E_D^{\textrm{max}}=\frac{1}{2}m_e
((v_F+2v_{\textrm{esc}})^2-v_F^2)$. When the DM is lighter (solid
curves), the cutoff is determined by the kinetic energy of the DM,
namely by $\mu_{eX} v_{\rm esc}^2/2$.

\begin{figure*}[ht!]
\begin{center}
\includegraphics[width=1\textwidth]{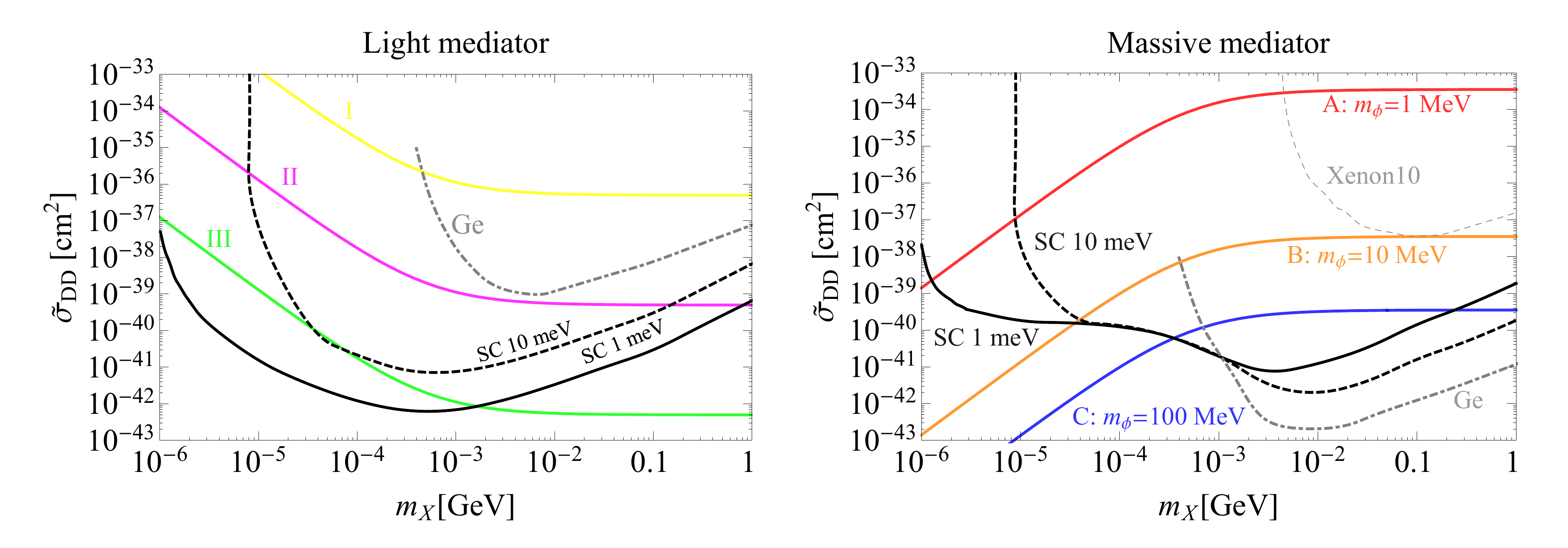}
\caption{
\label{fig:DDcomb}
%{\bf Left:} Upper bounds on direct detection cross section for light dark matter scattering off electrons, for very light mediators. Constraints arise from stellar cooling processes~\cite{An:2013yfc,An:2013yua}, bullet-cluster and halo shapes~\cite{Clowe:2003tk,Markevitch:2003at,Randall:2007ph,Rocha:2012jg,Peter:2012jh}, as well as kinetic decoupling during recombination epoch~\cite{McDermott:2010pa}.
{\bf Left:} Direct detection cross section for light dark matter scattering off electrons, for several benchmarks of light mediators. These are {\bf I:} $\alpha_X=10^{-15}, \alpha_e=10^{-12}$; {\bf II:} $\alpha_X=\alpha_e=10^{-15}$; and {\bf III:} $\alpha_X=10^{-15}, \alpha_e=10^{-18}$. These depicted parameters obey bounds from self-interactions and decoupling at recombination for $m_\phi\lsim {\rm eV}$, though stellar emission may place strong constraints for scalar mediators; see text for details.
{\bf Right:} Direct detection cross section between light dark matter and electrons, for several benchmarks of heavy mediators.
These are {\bf A:} $m_\phi=1$~MeV, $g_e=10^{-5}e$, $\alpha_X=0.1$; {\bf B:} $m_\phi=10$~MeV, $g_e=10^{-5}e$, $\alpha_X=0.1$; and {\bf C:} $m_\phi=100$~MeV, $g_e=10^{-4}e$, $\alpha_X=0.1$.
These depicted parameters obey all terrestrial and astrophysical constraints, though sub-MeV DM interacting with SM through a massive mediator may be strongly constrained by BBN; see text for details. The Xenon10 electron-ionization data bounds~\cite{Essig:2012yx} are plotted in thin dashed gray.
{\bf In both panels}, the black solid (dashed) curve depicts the sensitivity reach of the proposed superconducting detectors, for a detector sensitivity to recoil energies between 1~meV$-$1~eV (10~meV$-$10~eV), with a kg$\cdot$year of exposure. For comparison, the gray dot-dashed curve depicts the expected sensitivity utilizing electron ionization in a germanium target as obtained in Ref.~\cite{Essig:2011nj}.
}
\end{center}
\end{figure*}

%%%%%%%%%%%%%%%%%
\mysection{Results}
In Fig.~\ref{fig:DDcomb} we show the
$95\%$ expected sensitivity reach after one kg$\cdot$year exposure,
corresponding to the cross section required to obtain 3.6 signal
events~\cite{FeldmanCousins}. The left (right) panel corresponds to the light (heavy)
mediator regime, where we plot $\tilde \sigma_{\rm DD}^{\rm light}$
($\tilde \sigma_{\rm DD}^{\rm heavy}$) as a function of $m_X$. The
black solid [dashed] curve in both panels corresponds to a sensitivity to measured energies between 1~meV$-$1~eV [10~meV$-$10~eV]. For light mediators, the
scattering rate is sensitive to the lowest energy depositions,
resulting in a large improvement in reach when the detector
threshold is decreased. For massive mediators, the differential rate
peaks towards larger energies, though with a lower threshold there is more sensitivity to lighter particles.

%The next important question is what range of cross-sections in
%Eq.~\eqref{eq:DD} are consistent with astrophysical and terrestrial
%constraints on the couplings $\alpha_e$ and $\alpha_X$ of $\phi$ to
%electrons and DM.
For a sense of the size of the cross-sections in Eq.~\eqref{eq:DD}, we divide our discussion into light mediator and heavy mediator regimes. We begin with a light mediator $\phi$, which for illustration purposes we take to be a scalar.
%A bound on $\alpha_X$ is derived from DM self-interactions---the
%bullet-cluster~\cite{Clowe:2003tk,Markevitch:2003at,Randall:2007ph}
%along with recent simulations which reanalyze the constraints from
%halo shapes~\cite{Rocha:2012jg,Peter:2012jh}, limit the DM
%self-interacting cross section (at velocities
%$\gtrsim300\units{km/sec}$) to be $\sigma_T/m_X \lsim 1 \
%\units{cm^2/g}$, where we use the full expressions for
%(the classical regime of) $\sigma_T$ found {\it e.g.} in Ref.~\cite{Tulin:2012wi}.  The
%self-scattering constraint on $\sigma_T$ then places an upper bound
%on $\alpha_X$ for a given $m_\phi$ and $m_X$.
In the left panel of Fig.~\ref{fig:DDcomb} we
plot the direct detection cross section
$\tilde\sigma_{\rm DD}^{\rm light}$ [Eq.~\eqref{eq:DD}] for several benchmark points labeled {\bf I-III},
%for a variety of light mediator masses $m_\phi\lsim {\rm eV}$,
shown in solid colored curves.
As is evident, large direct detection cross sections can be obtained even for extremely small couplings due to the large enhancement factor in Eq.~\eqref{eq:DD}, that scales like four powers of the inverse of the momentum transfer in the detection process when the mediator is light. The presented benchmark points all obey DM self-interaction bounds~\cite{Clowe:2003tk,Markevitch:2003at,Randall:2007ph,Rocha:2012jg,Peter:2012jh} and also ensure that the DM remains out of kinetic equilibrium with the baryons up
through the time of recombination~\cite{McDermott:2010pa} for $m_\phi\lsim{\rm eV}$.  Stellar constraints are model-dependent (for example, whether a scalar or kinetically mixed vector mediator), and hence have not been factored in here; we note that for a kinetically mixed hidden photon, the strength of stellar constraints is lifted for the couplings shown in the plot since the combination $\sim g_X g_e$ or $m_\phi g_e$ is then bounded~\cite{An:2013yfc,An:2013yua} rather than just $g_e$.
Also note that the reach curves do not include any medium-dependent mediator mass, as this is model-dependent.  For example, in a metal, a kinetically mixed vector mediator would experience a large in-medium mass; such a mass becomes small in an insulating superfluid absorber like helium. %-3.  
We detail the medium- and model-dependence in a longer paper~\cite{future}.

Moving to heavy mediators, we focus on $m_\phi\gsim$~MeV. A plethora of constraints exists in the literature for this mass range, see {\it e.g.}~\cite{Bjorken:2009mm,Dreiner:2013mua,Dreiner:2013tja,Batell:2014mga} in the context of kinetically mixed hidden photons. In the right panel of Fig.~\ref{fig:DDcomb}, we select several benchmark points, labeled {\bf A-C}, that survive all terrestrial ({\em e.g.} beam dump) and stellar cooling constraints, %(depending on DM mass range as well)
and plot the resulting direct detection cross section of Eq.~\eqref{eq:DD}, $\tilde\sigma_{\rm DD}^{\rm heavy}$.  Large couplings to electrons $g_e \gsim 10^{-6}$ are possible despite stellar constraints due to trapping effects, and beam dump constraints may be evaded by decaying to additional particles in the dark sector. These statements hold regardless of the vector/scalar nature of the heavy mediator.
However, for values of $\alpha_X$ and $g_e$ as large as these benchmark points, DM and/or the mediator will be brought into thermal equilibrium with the SM plasma. The chief constraint on these models is thus BBN and Planck limits on the number of relativistic species in equilibrium (see {\em e.g.}~\cite{Boehm:2013jpa}).
The Planck constraints can be evaded; for instance coupling to $\gamma/e$ through the time that the DM becomes non-relativistic will act to reduce the effective number of neutrinos at CMB epoch.  On the other hand, during BBN, the helium fraction constrains the Hubble parameter, which is sensitive to all thermalized degrees of freedom.  DM must then be either a real scalar or heavier than a few hundred~keV in such simple models~\cite{Boehm:2013jpa}. It follows that part of the depicted curves of benchmarks {\bf A-C} in the low-mass region may not be viable; a detailed study of the viable parameter space is underway~\cite{future}. For completeness, we show the Xenon10 electron-ionization bounds~\cite{Essig:2012yx} in the thin gray dashed curve. (The Xenon10 bounds on light mediators are not depicted in the left panel of Fig.~\ref{fig:DDcomb} as they are orders of magnitude weaker than the parameter space shown.)

For comparison, we show the expected sensitivity using electron-ionization techniques with a germanium target as obtained in Ref.~\cite{Essig:2011nj}, translating their result into $\tilde \sigma_{\rm DD}$ of Eq.~\eqref{eq:DD}. These results are depicted by the dot-dashed gray curves in Fig.~\ref{fig:DDcomb} for both the light (left panel) and heavy (right panel) mediator cases. For heavy mediators and $m_X$ larger than a few hundred keV, our detection method is less sensitive than the projected one using germanium, while for lighter $m_X$, where electron ionization methods lose sensitivity, the superconducting devices win. (Indeed, this comparison between the detection methods is our main aim in presenting the right panel of Fig.~\ref{fig:DDcomb}.) In contrast, light mediators highlight the strength of our proposed detectors.
For DM masses above several hundred keV, superconducting detectors can out-perform electron ionization techniques by several orders of magnitude. For %super light
dark matter below the MeV scale, the proposed superconducting detectors are uniquely staged to detect super light sub-MeV viable models of dark matter.

In summary, we have proposed a new class of detectors that utilize superconductors to detect electron recoils from thermal DM as light as a keV. Given some improvement over current technology, such detectors may have sufficiently low noise rates to be sensitive to the required energy scale of meV electron recoils.  We have computed the DM scattering rates, taking into account Pauli blocking, and have shown that viable models may be detected.
We hope this proof of concept encourages the experimental community to pursue research and development towards the feasibility of such devices, probing detection of DM down to the keV scale.  We leave for future work the extended study of broader classes of DM models that may be detectible with these devices.

%%%%%%%%%%%%%%%%%%%
\mysection{Acknowledgments}
We thank Ehud Altman, Snir Gazit, Roni Ilan, Dan McKinsey, Dave Moore, Joel Moore and Zohar Ringel for useful discussions, and Jeremy Mardon for helpful correspondence and comments on the manuscript. We are especially grateful to John Clarke and Matt Pyle for critical conversations on viable detector designs. The work of YH is supported by the U.S. National Science Foundation under Grant No. PHY-1002399. YH is an Awardee of the Weizmann Institute of Science -- National Postdoctoral Award Program for Advancing Women in Science. YZ is
supported by ERC Grant BSMOXFORD No. 228169.  KZ is supported by the DoE under contract DE-AC02-05CH11231.

\end{document}